\documentclass[letterpaper,english,aps, nofootinbib, prd, twocolumn, lengthcheck, superscriptaddress]{revtex4-2}
\usepackage[T1]{fontenc}

\usepackage[colorlinks=true,linktocpage=true,linkcolor=blue,citecolor=blue]{hyperref}
\usepackage{verbatim}
\usepackage{bm}
\usepackage{amsmath}
\usepackage{amssymb}
\usepackage{graphicx}
\usepackage{slashed}
\usepackage{wrapfig}
\usepackage{bbm}
\usepackage[caption=false]{subfig}
\usepackage{verbatim}
\newcommand \beq{\begin{eqnarray}}
\newcommand \eeq{\end{eqnarray}}

\newcommand{\vp}{ {\bf p}}
\newcommand{\vq}{ {\bf q}}
\newcommand{\vk}{ {\bf k}}

\newcommand{\rp}{ {\rm p} }
\newcommand{\rma}{ {\rm a} }

\newcommand{\calC}{\mathcal{C}}
\newcommand{\calG}{\mathcal{G}}

\newcommand{\rmd}{\mathrm{d}}
\newcommand{\rmi}{\mathrm{i}}
\newcommand{\rme}{\mathrm{e}}

\begin{document}
\title{Thermal quarks and gluon propagators in two-color dense QCD}

\author{Toru Kojo}
\email{torujj@mail.ccnu.edu.cn}
\affiliation{Key Laboratory of Quark and Lepton Physics (MOE) and Institute of Particle Physics, Central China Normal University, Wuhan 430079, China}

\author{Daiki Suenaga}
\email{suenaga@rcnp.osaka-u.ac.jp}
\affiliation{Research Center for Nuclear Physics, Osaka University, Ibaraki, 567-0048, Japan}

\date{\today}

\begin{abstract}
We study Landau gauge gluon propagators in two-color QCD at finite quark chemical potential ($\mu_q$) and temperature ($T$). We include medium polarization effects at one-loop by quarks into massive gluon propagators,
and compared the analytic results with the available lattice data. 
We particularly focus on the high density phase of color-singlet diquark condensates whose critical temperature is $\sim 100$ MeV with weak dependence on $\mu_q$. 
At zero temperature the color singlet condensates protect the IR limit of electric and magnetic gluon propagators from the medium screening effects. 
At finite temperature, this behavior remains true for the magnetic sector, but the electric screening mass should be generated by thermal, and hence gapless, particles which are unbound from the diquark condensates. 
Treating thermal excitations as quasi-quarks, we found that the electric screening develops too fast compared to the lattice results. 
Beyond the critical temperature for diquark condensates the analytic results are consistent with the lattice results.

\end{abstract}

\maketitle

\section{Introduction}
\label{sec:intro}

Recently there have been growing interests on the dynamics at baryon density ranging from $\sim 5n_0$ to $\sim 40n_0$ in the context of neutron star physics \cite{Baym2017,Kojo:2020krb,Leonhardt,Rajan,Fukushima:2020cmk}. In this domain baryons are supposed to overlap and hence quarks and gluons should be natural degrees of freedom \cite{Masuda,kojo2015}, while the matter is likely to be strongly correlated as one can infer from the breakdown of the perturbative calculations around $\sim 40n_0$ \cite{Kurkela:2009gj,Kurkela:2014vha}. As the degrees of freedom are rather clear-cut, it is reasonable to expect the existence of some resummation with which strong $\alpha_s$ effects can be absorbed into the parameters of {\it quasi-particles}, e.g., effective mass, effective residues, so on, and after which residual interactions should become under control.

Along this line of thoughts, recently we began to study quasi-particle descriptions of quarks and gluons in very dense matter \cite{Suenaga:2019jjv,Song:2019qoh}. Our descriptions are based on the Landau gauge QCD which has been most extensively studied in 
functional methods \cite{vonSmekal:1997ern,Alkofer:2000wg,Alkofer:2006jf,Fischer:2008uz,Cyrol:2016tym}
and lattice gauge theories at 
zero baryon density \cite{Cucchieri:2011ig,Maas:2014xma}.
In this gauge the gluon propagators in the IR are tempered, and with a sufficiently large gluon mass the infrared divergences associated with the perturbative running coupling constant can be avoided. 
Especially in the decoupling solution, which has been favored in lattice calculations, a gluon acquires the mass, $m_g \sim 500$ MeV. These findings can be efficiently captured in the massive extension of the Yang-Mills theory in the Landau gauge \cite{Curci:1976bt,Tissier:2011ey,Reinosa:2017qtf,Reinosa:2013twa,Reinosa:2016iml,Pelaez:2014mxa}. We expect that the presence of this mass makes our Feynman graph calculations better organized, sharpening our questions on the genuinely non-perturbative effects. 

This point of view must be tested for finite density calculations. The two-color QCD is suitable for this purpose \cite{Kogut:2000ek,Kogut:1999iv}, as this system allows us to perform 
lattice Monte-Carlo simulations at finite chemical potential \cite{Boz:2019enj,Boz:2018crd,Boz:2013rca,Cotter:2012mb,Hands:2011ye,Hands:2010gd,Hands:2007uc,Hands:2006ve,Astrakhantsev:2020tdl,Bornyakov:2020kyz,Bornyakov:2017txe,Braguta:2016cpw,Braguta:2015zta,Braguta:2014gea,Braguta:2013loa,Astrakhantsev:2018uzd,Buividovich:2020gnl,Puhr:2016kzp,Wilhelm:2019fvp,Iida:2019rah,Muroya:2002ry,Makiyama:2015uwa}. 
The phase structure was analyzed by several analytic or continuum methods \cite{Sun:2007fc,He:2010nb,Kanazawa:2009ks,Brauner:2009gu,Andersen:2010vu,Andersen:2015sma,Contant:2019lwf,Strodthoff:2013cua,Strodthoff:2011tz} .
While at low density a dilute matter in two-color QCD is very different from the three-color QCD as baryons in the former are bosons, at high densities baryons overlap and the quark Fermi sea will be anyway established. It is this regime where we try to test our conjecture on the quasi-particle descriptions for the application to three-color QCD.

In the previous papers we have studied gluon propagators with \cite{Suenaga:2019jjv} and without \cite{Kojo:2014vja} gluon masses by including the medium quark loops  in the gluon polarizations. The calculations were performed at zero temperature in the presence of the color-singlet diquark condensates. At one loop, neither electric nor magnetic screening masses are generated; the electric sector is protected by the quark gaps, while the magnetic sector does not acquire the Meissner mass as the phase fluctuations of condensates do not couple to gauge fields\footnote{Similar situations have been discussed in two-flavor color superconductivity \cite{Rischke:2000qz,Rischke:2000cn}. }. Including the gluon mass tempers the impact of medium polarization effects both in electric and magnetic propagators, while the presence of diquark gaps substantially weakens the electric corrections. These two effects seem necessary to reproduce the lattice and not to spoil the systematics of computations. 
As for corrections beyond one-loop, the case without diquark gaps has been studied in a Dyson-Schwinger framework, 
and this work shows that electric propagators are over-suppressed compared to the lattice data \cite{Contant:2019lwf}. 
This suggests that the diquark gaps, among other non-perturbative effects, are indispensable to account for the lattice data.

Compared to the results based on pure perturbative gluons and quarks, the aforementioned quasi-particle picture substantially improves the consistency between the analytic results and the lattice's at zero temperature \cite{Suenaga:2019jjv}. 
In this paper we extend the analyses to the thermal medium.

\begin{figure}[tbp]
   \centering
   \vspace{-0.5cm}
  \includegraphics[scale=0.28]{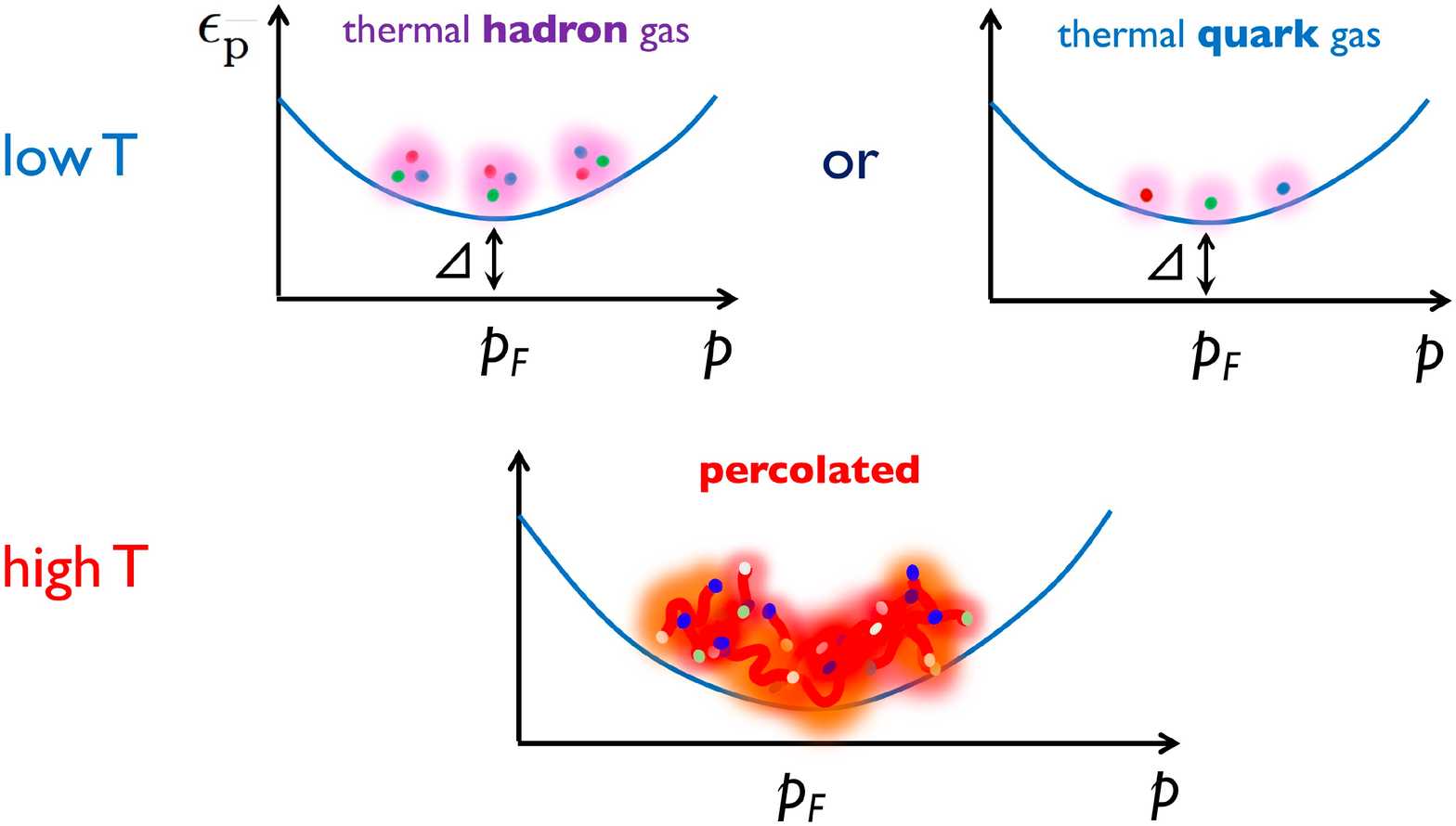} 
  \vspace{-0.5cm}
   \caption{Thermal excitations at low and high temperature. At low temperature a thermal gas is made by hadrons or quarks. At high enough temperature the gas becomes dense and get percolated. The percolation may happen within the superfluid phase if the phase space for low energy excitations is sufficiently large, e.g., at high density.}
   \label{fig:thermal_excitations}
\end{figure}

One of new questions arising at finite temperature is whether thermal excitations appear as quarks, or those excited quarks form color-singlet objects as in vacuum. For the schematic picture, see Fig.\ref{fig:thermal_excitations}.
Answering this question has the direct relevance to the quark-hadron continuity at finite temperature \cite{Kojo:2020ztt},  and to the quarkyonic matter conjecture which states that the bulk quark Fermi sea and baryonic structure near the Fermi surface \cite{McLerran:2007qj,McLerran:2018hbz,Pisarski:2018bct,Andronic:2009gj,Hidaka:2008yy,Duarte:2020kvi,Jeong:2019lhv,Kojo:2011fh,Ferrer:2012zq,McLerran:2008ua,Cao:2020byn,Zhao:2020dvu,Sen:2020peq}. 
Also, since the lattice data available so far have not reached temperatures less than $T \sim 40$ MeV,
it is important to prepare analytic results whose setup is close to the lattice's.

The smallest temperature reached on the lattice is $\simeq 44$ MeV. This temperature is not very low compared to the critical temperature of the diquark superfluidity,  $T_{\rm SF} \simeq $ 90-120 MeV, found on the lattice. Therefore we expect that the temperature corrections are not negligible.
Below in most cases we will use the Bardeen-Cooper-Schrieffer (BCS) formulas, which are valid at weak coupling \cite{Pisarski:1999bf}, as our baseline.
In this approximation the diquark gap at zero temperature is related to the critical temperature as\footnote{The critical temperature of the BCS is modified by the corrections such as Popov and Gor'kov-Melik-barkhudarov corrections even in the weak coupling limit, reducing the critical temperature by a factor $\simeq 2.2$. But these corrections are essentially modifications of effective interactions and reduce $\Delta$ in the same way. So the applicability of the BCS ratio $T_c \simeq 0.57 \Delta$ can be actually broader than that for the BCS estimate of each absolute value. See, e.g., Ref.\cite{Gorkov_modern} for nice summary.}
\beq
\Delta_{T=0} \simeq T_{\rm SF}/0.57 \simeq 158-211\, {\rm MeV}\,,
\label{eq:Del0}
\eeq
and the temperature dependence is
\beq
\Delta_T \simeq \Delta_{T=0} \, \big(1-T/T_{\rm SF} \big)^{1/2} \,.
\label{eq:DelT}
\eeq
These should be reasonable estimates for a chemical potential $\mu_q \gtrsim 1$ GeV where we expect the validity of weak coupling pictures. We simply assume its extrapolation to a lower $\mu_q$ to give a useful guide.

Provided $\Delta_{T=0}=200$ MeV, the gap at $T\simeq 44$ MeV is $\Delta_{T}\simeq 157$ MeV. If thermal excitations are quarks, the Bolzmann factor is $\sim \rme^{-\Delta_T/T} \sim \rme^{-157/44}\sim 0.03$. But this suppression factor is not small enough to dominate over the phase space factor for low energy quarks, $\sim p_F^2 \sqrt{\Delta_T T  \,} $, where $p_F$ is the quark Fermi momentum. Thermal quarks behave as gapless particles as they are already excited, and hence contribute to the electric screening, in the same way as quarks in a normal phase. This introduces non-negligible effects in the electric sector. Meanwhile, in the magnetic sector, no magnetic mass is generated. 

Another interesting possibility is that thermal excitations appear as hadrons, rather than individual thermal quarks. Then, at low temperature thermal corrections from them are more strongly suppressed than in the thermal quark case. This should continue until thermal quarks and gluons are liberated through the overlap of thermally excited hadrons. This liberation of colors is driven by entropic effects that compensate the Boltzmann factor, like in the case of Hagedorn gas \cite{Hagedorn:1965st,Polyakov:1978vu,Banks:1979fi,Hanada:2014noa}.
In this picture of deconfinement, the critical temperature decreases as density increases, as more phase space is available for low energy excitations and hence the entropy increases (provided that $\Delta_T$ is not sensitive to density or $\mu_q$).
With this picture in mind we examine the temperature dependence of the lattice data at high density.

This paper is organized as follows:
In Sec. \ref{sec:analytic} we present the one-loop expressions for the polarization tensors from gluonic loops and quark loops.
In Sec. \ref{sec:lattice}, after summarizing the setup used for the lattice simulations, we compare our one-loop results with the lattice's.
Sec. \ref{sec:summary} is devoted to summary and discussions.

%

\section{Analytic results}
\label{sec:analytic}

One-loop renormalized gluon polarization tensor in massive Yang-Mills (YM) theories has been computed in Ref.\cite{Tissier:2011ey}. The results read
\begin{eqnarray}
\Pi^R_{ {\rm YM} }(k) 
&=& \frac{g_s^2 K^2}{192\pi^2}\Bigg\{111{s}^{-1}-2{s}^{-2}+(2-{s}^2){\rm ln}({s}) \nonumber\\
&+& 2({s}^{-1}+1)^3({s}^2-10{s}+1){\rm ln}(1+{s}) \nonumber\\
&+& (4{s}^{-1}+1)^{3/2}({s}^2-20{s}+12) \nonumber\\
&\times & {\rm ln}\left(\frac{\sqrt{4+{s}}-\sqrt{{s}}}{\sqrt{4+{s}}+\sqrt{{s}}}\right)-({s} \leftrightarrow {\mu}_R^2/m_g^2)\Bigg\} \ , 
\label{PiGc} 
\end{eqnarray}
where $s = K^2/m_g^2$, with $K^2 = k_4^2+\vk^2$ being squared four-momenta in the Euclidean space, $m_g$ the gluon mass, $g_s$ the coupling constant. The renormalization scale $\mu_R$ is chosen to be $\simeq 1$ GeV, where the overall size of our gluon propagators is set to reproduce the lattice propagator at $\mu_R$.
When we consider finite temperature gluon propagators, we neglect thermal gluon loops which are suppressed at temperature ($T\lesssim 200$ MeV) substantially smaller than the energy of massive gluons ($m_g\sim 500$ MeV).

In medium, the polarization effects caused by quarks are computed in the Nambu-Gor'kov formalism assuming the presence of a momentum independent gap $\Delta_T$.
It is convenient to decompose the quark polarization tensor into the vacuum and medium pieces. Together with the vacuum counter terms, the renormalized polarization function is
\beq
\Pi_{E,M}^{q, R} 
= \Pi_{E,M}^{q}\big|_{\rm bare } - k^2 \delta_{Z_g} 
=\Pi^{q, R}_{\rm vac} + \delta \Pi^q_{E,M} \,,
\eeq
where the renormalized vacuum polarization and the medium corrections are
\beq
\Pi_{\rm vac}^{q, R} &=& \Pi^{q}_{\rm vac} \big|_{\rm bare } - k^2 \delta_{Z_g} \,,
\nonumber \\
\delta \Pi^q_{E,M} &=& \Pi_{E,M}^{q} \big|_{\rm bare } - \Pi_{\rm vac}^q \big|_{\rm bare } \,.
\eeq
The vacuum part is treated as in usual perturbation theories. The second term is the bare medium and vacuum contributions both of which are UV divergent, and the subtraction leads to the UV finite expression. In the usual method to pick up poles of $p_4$-integrals, however, the implicit regularization of spatial momenta\footnote{We take $|\vp|\rightarrow \infty$ only after picking up the residue.} breaks the gauge invariance \cite{Kojo:2014vja}. The artifacts automatically cancel if we use the same quark propagators for the medium and vacuum, but with such conditions we would fail to capture the relevant physics associated with changes in quasi-particles. Fortunately the Ward identity allows us to identify the gauge variant artifacts, so one can use gauge variant counter terms to cancel them\footnote{ 
The procedure here uniquely specifies the gauge variant counter terms for a given external momentum $k$. But in principle there are still possibilities that the dependence on the gauge variant regularization would enter gauge invariant quantities as finite terms. There are, at least to us, no obvious way to identify such contributions. We leave this problem for the future.
}. 
Including such procedure, the expression becomes \cite{Suenaga:2019jjv}
\beq
 \delta \Pi_{E,M}^q = \delta_{\Delta S} \Pi^q_{E,M} + \delta_{\Delta \mu_q} \Pi^q_{E,M} \,.
 \label{eq:delPi}
\eeq
The first term takes care of changes in quark propagators, and is computed in the dimensional regularization,
\beq
&& \delta_{\Delta S} \Pi^{} (k) \big|_{ {\rm dim\, reg} }  = - K^2\frac{g_s^2}{\, 2\pi^2 \,} \nonumber \\
&& \times \int_0^1 dx\, x(1-x) {\rm ln} \frac{ ( \tilde{M}_q )^2 + x(1-x) K^2 }{\, (M^{ {\rm vac} }_q)^2 + x(1-x) K^2 \,}\,,
\eeq
with $\tilde{M}_q = \sqrt{\,   \Delta_T^2 + M_q^2 \,}$. Here $M_q^{\rm vac}$ is the constituent quark mass in vacuum whose value is set to $M_q^{\rm vac} =0.3$ GeV, while $M_q$ is the mass at high density which should be close to the current quark mass. We set $M_q=0.1$ GeV considering the large pion mass, $m_\pi \sim 0.7$ GeV. The form of $\tilde{M}_q$ is uniquely fixed to cancel the artifacts in the second term in Eq.(\ref{eq:delPi}). The expression is given by ($\int_{\vq} \equiv \int \rmd^3 \vq /(2\pi)^3$)
%
\begin{widetext}
\beq
\delta_{ \Delta \mu_q} \Pi^q_{E,M}(k) \big|_{ {\rm 3d\, reg} } 
= \Pi^q_{E,M}(k) \big|_{ {\rm bare} } - \Pi^q_{E,M}(k) \big|^{M_q \rightarrow \tilde{M}_q;\, \mu_q \rightarrow 0 ;\, \Delta \rightarrow 0}_{ {\rm bare} } \,,
\eeq
where
\beq
\Pi^q_{E,M}(k) \big|_{ {\rm bare} } 
= g_s^2  \sum_{ s, s' = \rp, \rma } \int_{ \vq }  {\cal K}_{E,M}^{ss'} (\vq_+, \vq_-) 
	\bigg[\,  {\cal C}_{E,M}^{ss'} ( \vq_+, \vq_- ) \,  {\cal G}_{ss'} (q_+,q_-) + \tilde{\calC}_{E,M}^{ss'} ( \vq_+, \vq_- ) \,  \tilde{\calG}_{ss'} (q_+,q_-) \,\bigg] \,.
\eeq
\end{widetext}
%
Here $\vq_\pm = \vq \pm \vk/2$. Below we write $E_{q_\pm} = \sqrt{ M_q^2 + \vq_\pm^2 }$ and the excitation energies for quasi-particles and quasi-antiparticles as $\epsilon_{\rp}^\pm = \sqrt{ (E_{q_\pm} - \mu_q)^2 + \Delta_T^2}$ and $\epsilon_{\rma}^\pm = \sqrt{ (E_{q_\pm} + \mu_q)^2 + \Delta_T^2}$, respectively.

The explicit forms of these factors are as follows: 
the kinematic factors are
\begin{eqnarray}
&& {\cal K}^{ \rp \rp}_{E} =  {\cal K}^{ \rma \rma}_{E}  = 1 + \frac{\, \vec{q}^2 - \vk^2/4 + M_q^2 \,}{ E_{q_+}E_{q_-} }  \ , 
\nonumber \\
&& {\cal K}^{ \rp \rma }_{E} =  {\cal K}^{ \rma \rp}_{E}  =  1 - \frac{\, \vec{q}^2 - \vk^2/4 + M_q^2 \,}{ E_{q_+}E_{q_-} } \ , 
\nonumber \\
&& {\cal K}^{ \rp \rp}_{M} =  {\cal K}^{ \rma \rma}_{M}  = -1 + \frac{\, ( |\vec{q}|\cos\theta)^2 - \vk^2/4 + M_q^2 \,}{ E_{q_+}E_{q_-}}  \ ,
\nonumber \\
&& {\cal K}^{ \rp \rma}_{M} =  {\cal K}^{ \rma \rp}_{M}  = -1 - \frac{\, ( |\vec{q}|\cos\theta)^2 - \vk^2/4 + M_q^2 \,}{ E_{q_+}E_{q_-}}  \ ,
\end{eqnarray}
where $\hat{k} = \vk/|\vk|$ and $\cos\theta$ is the angle between $\vq$ and $\vk$;
the coherence factors are
\begin{eqnarray}
{\cal C}^{ \rp \rp}_{E,M} &=& \frac{1}{2} \left( 1 - \frac{\, ( E_{q_+} - \mu_q  )(  E_{q_-}  - \mu_q ) \pm |\Delta_T |^2 \,}{ \epsilon_\rp^+ \epsilon_\rp^-  } \right) \,, \nonumber\\
{\cal C}^{ \rma \rma}_{E,M} &=& \frac{1}{2} \left( 1 - \frac{\, ( E_{q_+} +  \mu_q )( E_{q_-} + \mu_q ) \pm |\Delta_T |^2 \,}{ \epsilon_\rma^+ \epsilon_\rma^- } \right) \,, \nonumber\\
{\cal C}^{ \rp \rma}_{E,M} &=& \frac{1}{2} \left( 1 + \frac{\, ( E_{q_+} -  \mu_q )( E_{q_-} + \mu_q ) \mp |\Delta_T |^2 \,}{ \epsilon_\rp^+ \epsilon_\rma^- } \right) \,, 
\end{eqnarray}
and 
\begin{eqnarray}
\tilde{\calC}^{ \rp \rp}_{E,M} &=& \frac{1}{2} \left( 1 + \frac{\, ( E_{q_+} - \mu_q  )(  E_{q_-}  - \mu_q ) \pm |\Delta_T |^2 \,}{ \epsilon_\rp^+ \epsilon_\rp^-  } \right) \,, \nonumber\\
\tilde{\calC}^{ \rma \rma}_{E,M} &=& \frac{1}{2} \left( 1 + \frac{\, ( E_{q_+} +  \mu_q )( E_{q_-} + \mu_q ) \pm |\Delta_T |^2 \,}{ \epsilon_\rma^+ \epsilon_\rma^- } \right) \,, \nonumber\\
\tilde{\calC}^{ \rp \rma}_{E,M} &=& \frac{1}{2} \left( 1 - \frac{\, ( E_{q_+} -  \mu_q )( E_{q_-} + \mu_q ) \mp |\Delta_T |^2 \,}{ \epsilon_\rp^+ \epsilon_\rma^- } \right) \,, 
\end{eqnarray}
where ${\cal C}^{ \rp \rma}_{E, M} (q_+, q_-) = {\cal C}^{ \rma \rp}_{E,M} (q_-, q_+)$ and $\tilde{\calC}^{ \rp \rma}_{E, M} (q_+, q_-) = \tilde{\calC}^{ \rma \rp}_{E,M} (q_-, q_+)$; 
finally the propagator part is
\beq
{\cal G}_{ss'} (q_+,q_-) &=&  \frac{1}{\, 2 \,} \big(\, 1 -n(\epsilon_s^+)  -n(\epsilon_{s'}^-) \, \big) 
\nonumber \\
&& \hspace{-1cm} 
\times \bigg( \frac{1}{\, \rmi k_4 + \epsilon_s^+ + \epsilon_{s'}^- \,} 
+ \frac{1}{\, - \rmi k_4 + \epsilon_s^+ + \epsilon_{s'}^- \,} \bigg) \,,
\eeq
and 
\beq
\tilde{\calG}_{ss'} (q_+,q_-) &=& - \frac{1}{\, 2 \,} \big(\, n(\epsilon_s^+)  - n(\epsilon_{s'}^-) \, \big) 
\nonumber \\
&& \hspace{-1cm} 
\times \bigg( \frac{1}{\, \rmi k_4 + \epsilon_s^+ - \epsilon_{s'}^- \,} 
+ \frac{1}{\, - \rmi k_4 + \epsilon_s^+ - \epsilon_{s'}^- \,} \bigg) \,.
\eeq
The function $n(x)$ is the Fermi-Dirac distribution $n(x)=1/( \rme^{ \beta(x-\mu) }+1 )$.
Below we focus on the static behaviors of the gluon propagators at $k_4=0$ 
where the results are most sensitive to the nonperturbative effects.
Table. \ref{tab:medium} summarizes the coherence and kinematical factors for electric and magnetic gluons, and the factors from the propagators.

\begin{table}[bht]
\vspace{0.5cm}
\hspace{2.5cm}
\begin{tabular}{|r| c|c || c|c || c|}
\hline 
~~~~~~&~${\cal C}_E$~&~${\cal K}_E$~&~${\cal C}_M$~&~${\cal K}_M$~&~${\cal G} ( |\vq|=p_F)$~ \\ 
\hline 
~~pp~~~&~$\sim \vq \cdot \vk$~&~2~&~$\left( \frac{\Delta_T}{\epsilon_{ {\rm p} } (q) } \right)^2$~&~$\left( \frac{ \vq}{E_q} \sin \theta \right)^2$~&$\sim \frac{1}{\, \vq \cdot \vk + \Delta_T^2 \,}$ \\
~~aa~~~&~$\sim \vq \cdot \vk$~&~2~&~$\left( \frac{\Delta_T}{\epsilon_{ {\rm a} } (q) } \right)^2$~&~$\left( \frac{ \vq}{E_q} \sin \theta \right)^2$~&$\sim \frac{1}{\, p_F \,}$  \\
~~pa~~~&~finite~~&~$ \sim \vk^2$~~&~finite~~&~~$-2$~~&$\sim \frac{1}{\, p_F \,}$  \\ 
\hline
\end{tabular}
\caption{\footnotesize{  The coherence and kinematical factors for electric and magnetic gluons, and the factors from the propagators at $|\vq|=p_F$ where $p_F$ is the quark Fermi momentum such that $E(p_F)=\mu_q$.
 } }
\label{tab:medium}
\end{table}

The zero temperature limit was discussed in the previous papers \cite{Suenaga:2019jjv,Kojo:2014vja};
the electric part is dominated by the particle-hole contributions for any phases, as one can see from the kinematic factor. Then, in the case of color-singlet diquark condensed phases, the coherence factor vanishes due to the quark gap, while the gaps also introduce the infrared cutoff\footnote{If $\Delta_T =0$, the $\calC_{\rp \rp} \calG_{\rp \rp}$ becomes finite in the static limit, producing the Debye mass.} $\sim \Delta_T$ in the propagator factor $\calG_{\rp \rp}$. 
These facts together lead to vanishing electric polarization for $\vk \rightarrow 0$. 
On the other hand, the magnetic contributions come from everywhere, from $(\rp \rp)$, $(\rp \rma)$, and $(\rma \rma)$, which are correlated through the gauge invariance; for quark propagators in normal or color-singlet diquark condensed phases, the all contributions are assembled to cancel, leaving the vanishing magnetic contributions.

At finite temperature, the magnetic sector for the static limit is unchanged while the electric sector is no longer protected by the quark gap. There are thermally excited quarks which can be easily perturbed by external fields; these thermal quarks are gapless. Moreover they are not arranged into color singlet objects. Accordingly, the static limit ($k_0=0$, $|\vk | \rightarrow 0$) yields
\beq
&&\tilde{\calG}_{\rp \rp} 
~\rightarrow~
	 - \frac{\, \partial n(\epsilon_\rp)  \,}{\, \partial \epsilon_\rp \,}
= \frac{1}{\, T \,}  \frac{\,  \rme^{ \epsilon_\rp /T} \,}{\, \big( \rme^{ \epsilon_\rp /T} + 1 \big)^2 \,}  
~\sim~ \frac{\,  \rme^{-\epsilon_\rp/T} \,}{\, T \,} \,,
\nonumber \\
&&\tilde{\calC}_E^{\rp \rp} ~\rightarrow~ 1 \,.
\eeq
Here the coherence factor is $1$ like in a normal phase.
This contribution adds the Debye mass to the electric sector. This correction is very sensitive to the size of gap which controls the abundance of thermal quarks. At low temperature its size is $\sim  \rme^{-\Delta_T/T}/T$, exponentially suppressed. As temperature approaches the critical temperature of diquark condensations, $\Delta_T$ approaches zero; 
here the factor $\tilde{\calG}_{\rp \rp}$ is no longer exponentially suppressed and becomes $\sim 1/T$, and combining it with the phase space factor $\sim T p_F^2$ and coupling constants leads to the Debye mass of $m_D^2 \sim g_s^2 T^{-1} (Tp_F^2) \sim g_s^2 p^2_F$. 

\section{Comparison with the lattice data}
\label{sec:lattice}

\begin{figure}[tbph]
   \centering
   \vspace{-0.5cm}
  \includegraphics[scale=0.28]{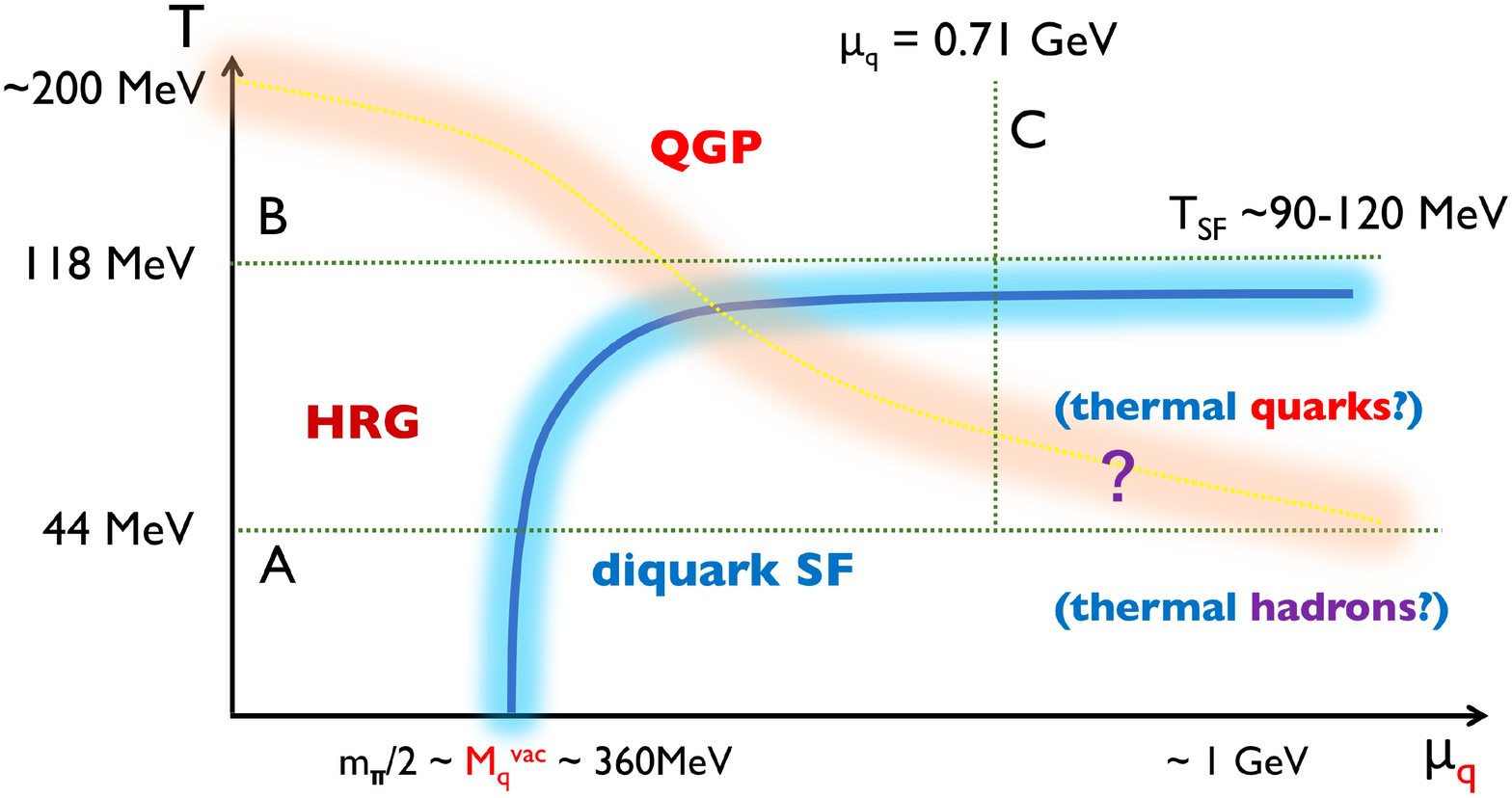} 
  \vspace{-0.5cm}
   \caption{A schematic phase diagram. The lines A, B, C indicate the domains analyzed in Sec.\ref{sec:lattice}A-C. See also Fig.10 in Ref.\cite{Cotter:2012mb} and Fig.2 in Ref.\cite{Iida:2019rah} based on the Polyakov loop.}
   \label{fig:phase}
\end{figure}

As in our previous study  \cite{Suenaga:2019jjv}, we will use the data in Ref.\cite{Boz:2018crd} based on
the gauge configurations of unimproved Wilson gauge action with 2 flavors of unimproved Wilson quarks.
The pion mass is $m_\pi = 717(25)$ MeV, so the onset chemical potential of the baryon density 
is $\mu_c =m_\pi/2 \simeq 358$ MeV\footnote{The impact of explicit symmetry breaking was examined in details by chiral effective models, see Ref.\cite{Andersen:2015sma}. }.
As the explicit chiral symmetry breaking is so large, we regard pions as two constituent quarks in the same way as $\rho$-mesons or other mesons.
The critical temperature of deconfinement, $T_D$, is defined by the Polyakov loop and $T_D = 217(23)$ MeV.

Our previous paper has compared the one-loop results with the lattice data for $\beta = 1.9$ and $N_t \times N_s^3 = 24 \times 16^3$ with the inverse lattice spacing $a^{-1} \simeq 1.06$ GeV, the spatial size $L_s \simeq 2.98$ fm, and the temporal size $L_t \simeq 4.46$ fm. While the data for this $L_t$ was presented as the zero temperature result \cite{Boz:2018crd}, this corresponds to $T \simeq 44$ MeV if we interpret $1/L_t$ as temperature. At zero density the result at $L_t > L_s$ is often interpreted as the zero temperature result as the finite volume makes the lowest momentum $\sim 1/L_s$ and hence lifts up the energy. But at finite density there is the Fermi sea and large momentum states do not necessarily mean high energy states\footnote{We acknowledge Profs. A. Maas and J. Skullerud for instructions on this point.}. Following Ref.\cite{Boz:2019enj} we literally take $1/L_t$ as temperature. This temperature corrections were not taken into account in our previous analyses and we shall include these corrections.

In this paper we compare our finite temperature expression with the lattice data. The data set for $\beta=2.1$ and $N_s =16$ with the inverse lattice spacing $a^{-1}=1.41$ GeV covers the wide range of temperatures, and this data set will be used in this work. The spatial size is $L_s \simeq 2.21$ fm and the temporal size covers the temperature range from $T\simeq 44$ MeV ($N_t=32$) to $\simeq 353$ MeV ($N_t=4$). See Appendix B of Ref.\cite{Boz:2018crd} where details of the simulation setup are summarized.

In order to describe the finite temperature, it is crucial to know the size and temperature dependence of the gap. As a baseline we simply assume the BCS expressions in Eqs.(\ref{eq:Del0}) and (\ref{eq:DelT}).
The critical temperature for the diquark condensation, $T_{\rm SF}$ is $\simeq$ 90-120 MeV for $\mu_q =705$ MeV, and at high density it is not very sensitive to changes in $\mu_q$. Thus throughout our analyses we take
\beq
\Delta_{T=0} = 200\, {\rm MeV} \,,
\eeq
for the zero temperature gap, and the corresponding critical temperature is
\beq
T_{\rm SF} = 114\, {\rm MeV} \,.
\eeq
Below we will substitute these expressions unless otherwise stated.

For comparisons of our results with the lattice data, we need to multiply an overall factor as the latter results were not renormalized. We use the expression
\beq
D^{\rm lattice}_{\rm E,M} (k ) = Z_{\rm lattice} \, D^{\rm analytic}_{E,M} (k) \,,
\eeq
and chose $Z_{\rm lattice} =3.0$ throughout. We use the lattice results in Ref.\cite{Boz:2018crd}; the bottom panels in Fig.11;  top panels in Fig.16; and top panels in Fig.19.
The domains we study in Sec.\ref{sec:lowT}-\ref{sec:highT} are summarized in Fig.\ref{fig:phase}, together with a schematic phase diagram.

\begin{figure}[htbp]
   \centering
  \hspace{-0.9cm} \includegraphics[scale=0.8]{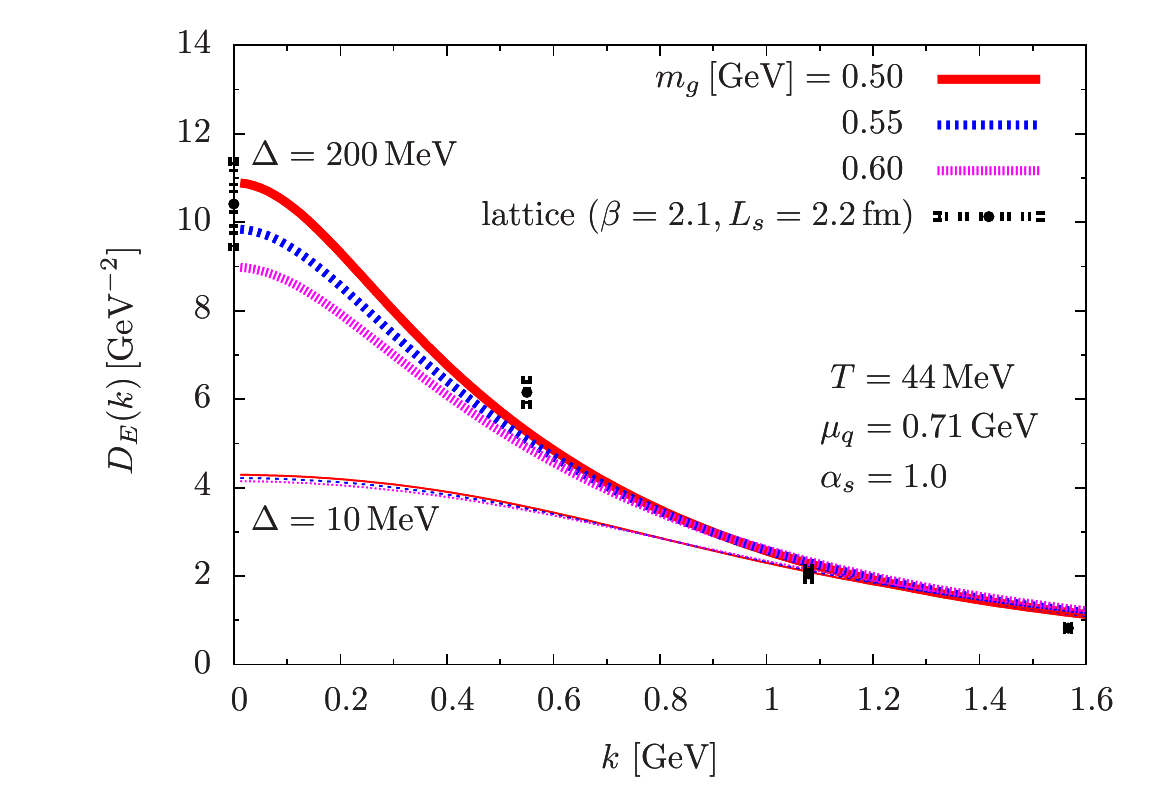} 
  \\
   \centering
  \hspace{-0.9cm} \includegraphics[scale=0.8]{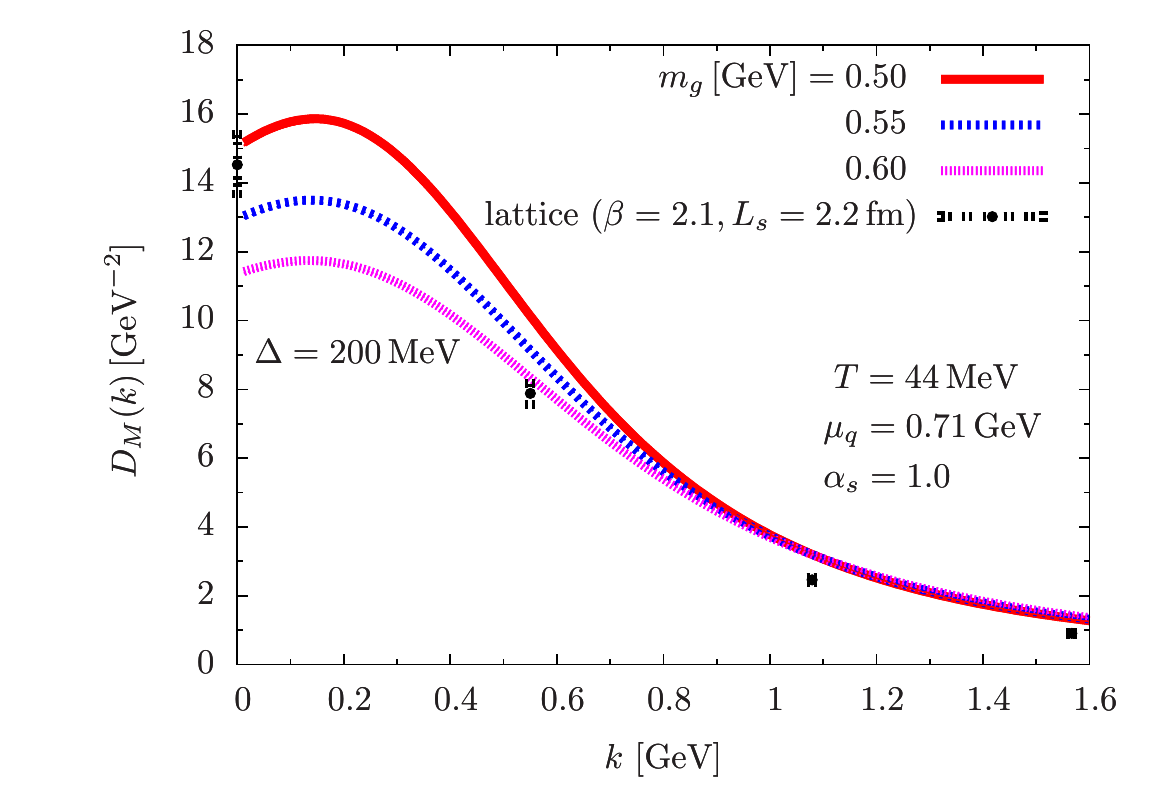} 
   \caption{ The static propagators at $\mu_q=0.71$ GeV and $T\simeq 44$ MeV, for $\alpha_s=1.0$. The upper (lower) panel is for electric (magnetic) propagators.  
   For the electric propagators we compare the $\Delta_{T=0} =$ 200 MeV and 10 MeV, while in the magnetic sector the dependence on $\Delta_T$ is not visible for the range 10-200 MeV.}
   \label{fig:muvaryT44_mg_vary}
\end{figure}

\subsection{The $\mu_q$-dependence at $T\simeq 44$ MeV}
\label{sec:lowT}

\begin{figure}[htbp]
   \centering
  \hspace{-0.9cm} \includegraphics[scale=0.8]{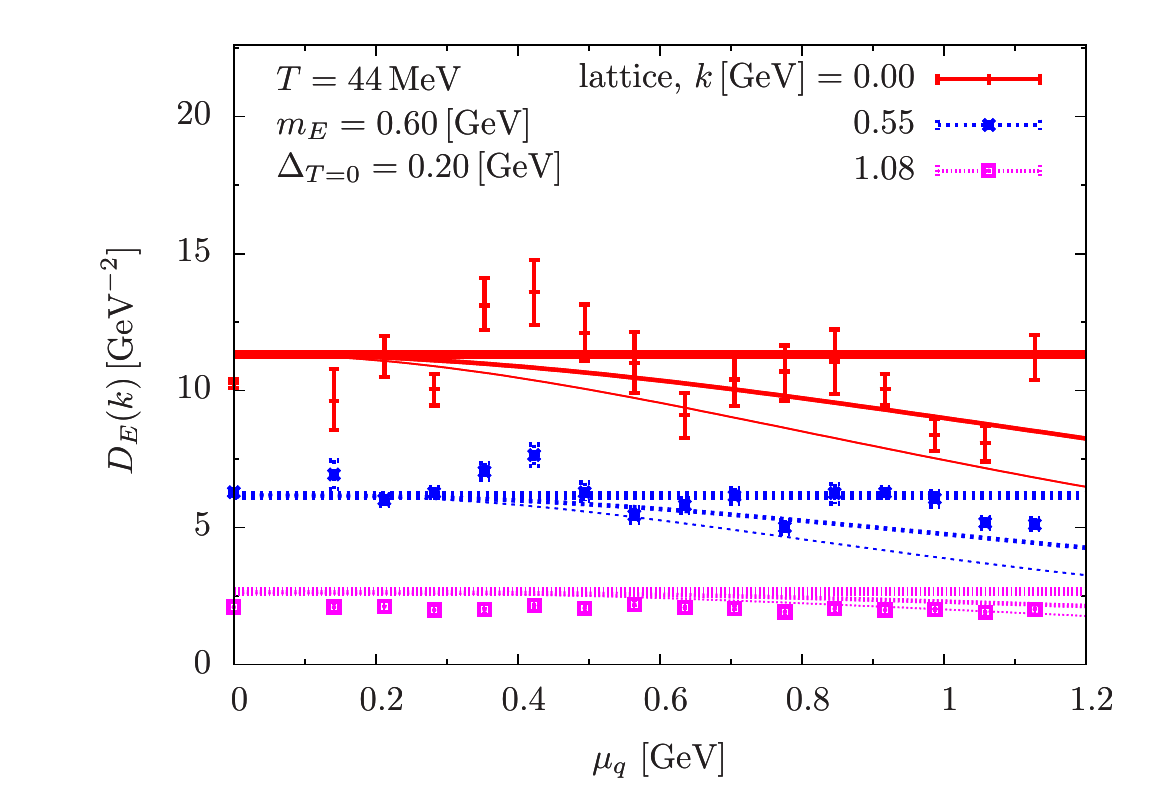} 
  \\
   \centering
  \hspace{-0.9cm} \includegraphics[scale=0.8]{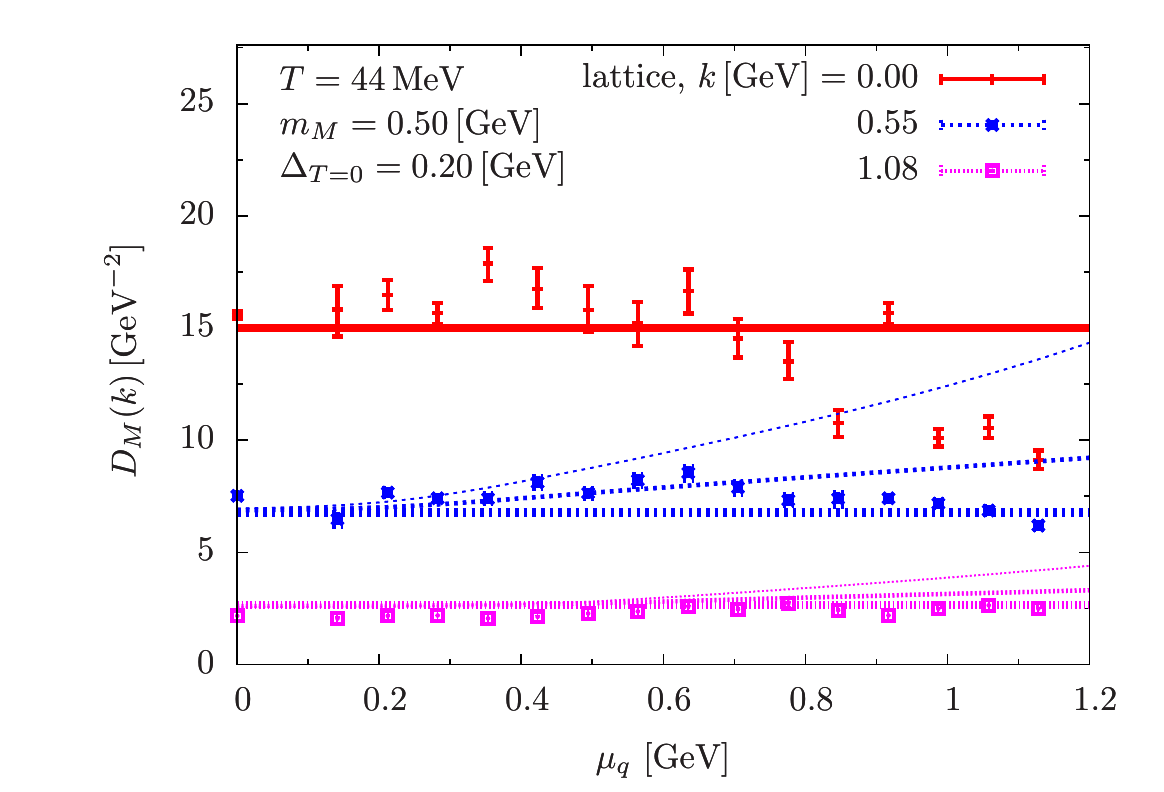} 
   \caption{The $\mu_q$-dependence of the gluon propagators at spatial momenta $|\vk|=0.0,\, 0.55,\, 1.08$ GeV and at temperature $T\simeq 44$ MeV. The upper (lower) panel is for electric (magnetic) propagators. The plots with error bars correspond to the lattice data. From bold to thin lines, they correspond to the analytic results at $\alpha_s=$ 0.0, 0.5, 1.0, and the $\Delta_{T=0} =0.2$ GeV and $\Delta_{T=44} \simeq 157$ MeV. }
   \label{fig:muvaryT44_fixed_mom}
\end{figure}

We first consider the low temperature case. For this case the dependence of gluon propagators on $m_g$, $\Delta_{T=0}$, $\alpha_s$, and momenta were studied in detail in Ref.\cite{Suenaga:2019jjv}, so in this paper we limit ourselves to fewer cases.


To get ideas about the overall impact of parameters and the tendency of electric and magnetic propagators, in Fig.\ref{fig:muvaryT44_mg_vary} we first show the $m_g$- and $\Delta_{T=0}$-dependence of gluon propagators (see also Ref.\cite{Suenaga:2019jjv}). The $\alpha_s$ is fixed to $ 1.0$. We note that there is disparity in the magnitudes of electric and magnetic propagators already in vacuum due to the fact that the lattice setup is anisotropic in temporal and spatial directions. 
Keeping this in mind, we vary $m_g$ by $\pm 10\%$ which changes the magnitudes of propagators by $\sim 10\%$ in the IR limit. 
The $\Delta_{T=0}$-dependence is very large in the electric sector, but in the magnetic sector such dependence is not visible. In medium the analytic results show that the magnetic propagators have the paramagnetic enhancement at finite momenta compared to the vacuum case. It is difficult to see this enhancement in the lattice results perhaps because the second smallest momentum is too large, $|\vk|\simeq 0.55$ GeV.

Below we examine the $\mu_q$-dependence of static gluon propagators at three momentum data points.
Shown in Fig.\ref{fig:muvaryT44_fixed_mom} is the $\mu_q$-dependence of the electric (upper) and magnetic (lower) gluon propagators $D_{E,M} (k)$ at momenta $|\vk| =0.0, 0.55, 1.08$ GeV. The lattice data are shown with error bars and the corresponding analytic results are shown with three lines; from bold to thin lines, we show the results for $\alpha_s=0.0, 0.5, 1.0$. For the electric and magnetic masses (tree level), we chose $m_E=0.6$ GeV and $m_M = 0.5$ GeV, respectively to get good overall fit. 

The overall behaviors are that in both sectors the gluon propagators are largely insensitive to $\mu_q$, except $\mu_q \gtrsim 0.8$ GeV where the propagator at $|\vk| =0$ starts to get suppressed gradually. It is not easy which $\alpha_s=$ 0.0 or 0.5 should be regarded as the better choice, but it seems that the choice $\alpha_s$ = 1.0 leads to too large changes especially at momenta $|\vk| = 0.55 $ GeV; in the electric sector due to the thermal Debye screening, and in the magnetic sector due to the paramagnetic enhancement.

Here we mention briefly the trend seen in the magnetic sector for $\mu_q \gtrsim 0.8 $ GeV where the magnetic propagators seem to decrease for increasing $\mu_q$. 
This trend cannot be explained in our one-loop calculations.
We note that in this domain of $\mu_q$ and $T$ the Polyakov loop starts to grow \cite{Cotter:2012mb}.
So we guess that the trend for $\mu_q \gtrsim 0.8 $ GeV in Fig.\ref{fig:muvaryT44_fixed_mom} originates from the beyond one-loop effects; presumably we need to address the modification of gluon mass at the nonperturbative level.

\subsection{The $\mu_q$-dependence at $T\simeq 118$ MeV}
\label{sec:highT}

\begin{figure}[htbp]
   \centering
   \hspace{-0.9cm} \includegraphics[scale=0.8]{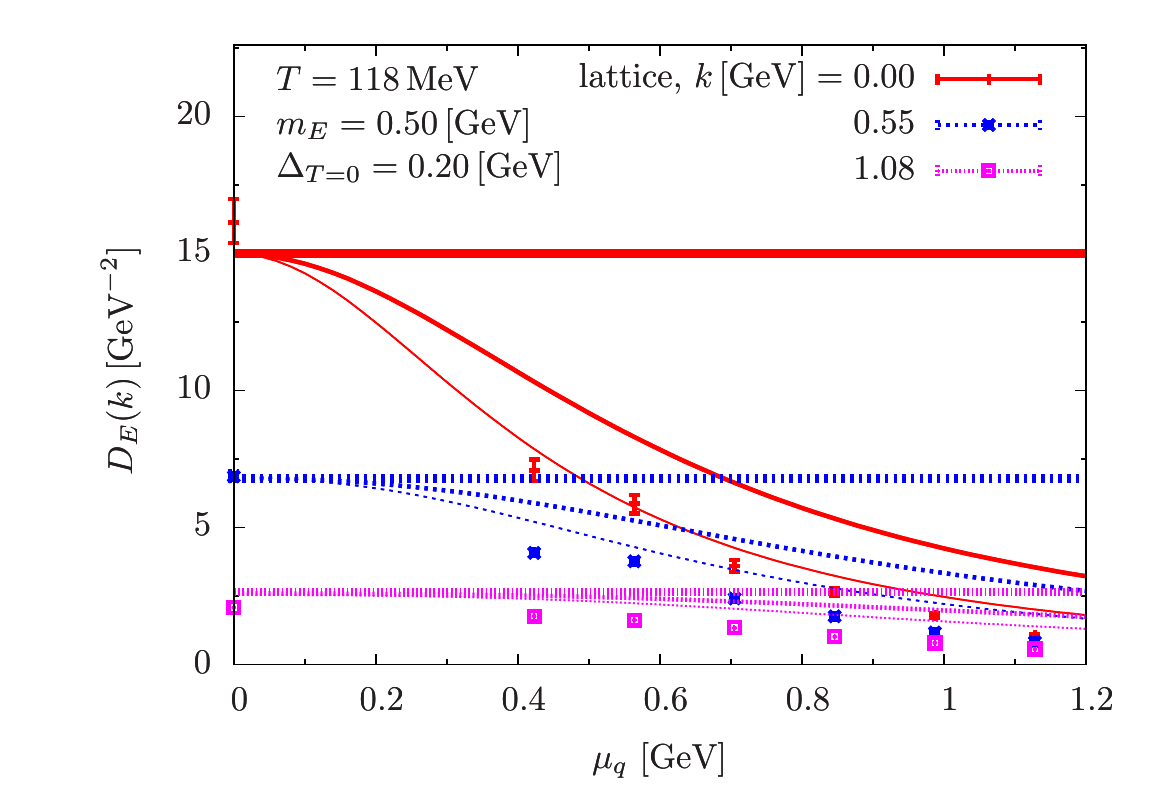} 
     \\
   \centering
   \hspace{-0.9cm} \includegraphics[scale=0.8]{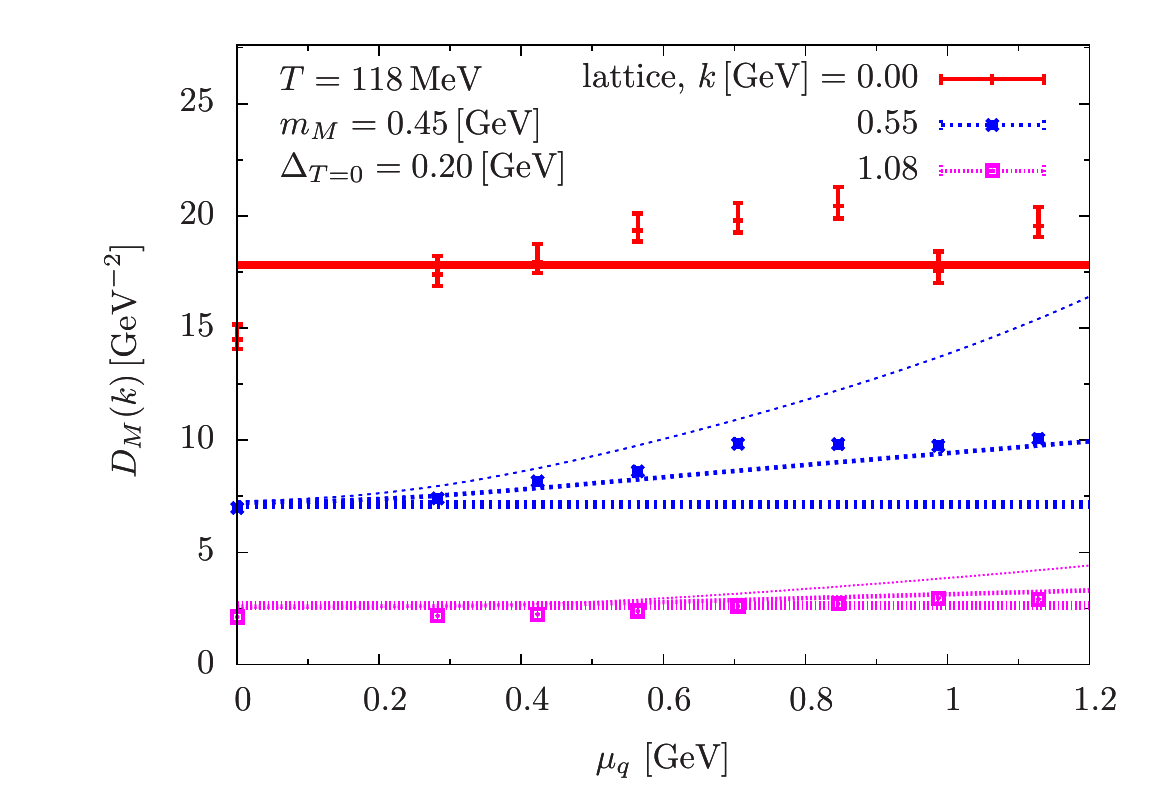} 
   \caption{The same as Fig.\ref{fig:muvaryT44_fixed_mom} except that the temperature is $T\simeq 118$ MeV and $m_E$ and $m_M$ change slightly.
    The temperature is higher than $T_{\rm SF}$, so $\Delta_{T=118} =0$.
   }
   \label{fig:muvaryT118_fixed_mom}
\end{figure}

Next we turn into higher temperature, $T\simeq 118$ MeV, and examine the $\mu_q$-dependence of the gluon propagators (Fig.\ref{fig:muvaryT118_fixed_mom}). 
We reduce the gluon mass slightly, $\sim 10\%$, instead of manifestly computing the gluon loop at finite temperature. At this temperature the diquark condensates melt or are vanishingly small. Also, the lattice results show that the Polyakov loop is large, meaning the abundance of thermal quarks. Therefore we may expect that the lattice data can be explained by computations based on deconfined thermal quarks.

The comparison between analytic results and the lattice results supports this picture. The electric propagators are drastically reduced by the Debye screening as $\mu_q$ increases, and the trend in the lattice data are reproduced for all spatial momenta. As for the magnetic sector, we have not fully understood the growth of the propagator at $|\vk|=0$ for increasing $\mu_q$, but the growing behavior at $|\vk| = 0.55$ and $1.08$ GeV is reasonably consistent with the paramagnetic enhancement in analytic computations. 

As for the $\alpha_s$-dependence, although we have not tried a fine-tuning, the results for the range of $\alpha_s = $ 0.5-1.0 seem to give reasonable descriptions for the $\mu_q$-dependence.

\subsection{The $T$-dependence at $\mu_q =0.71$ GeV}
\label{sec:highT}

%
\begin{figure}[htbp]
   \centering
   \hspace{-0.9cm} \includegraphics[scale=0.8]{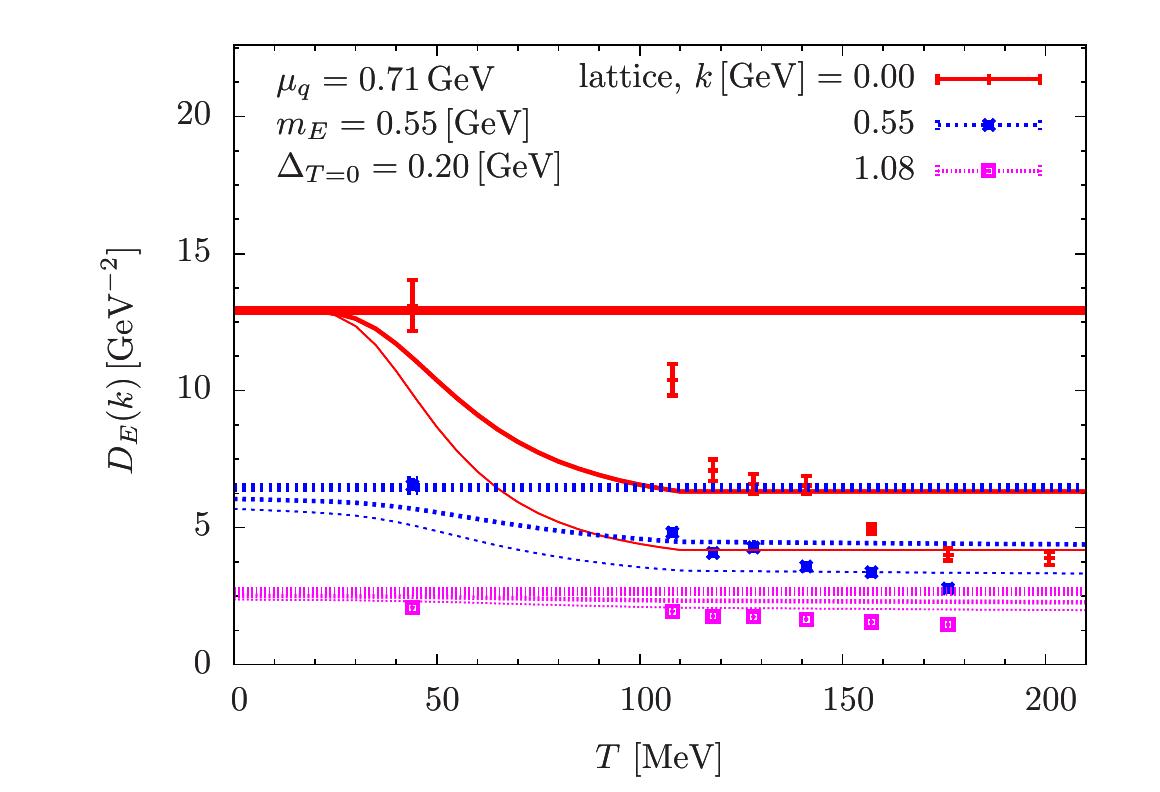} 
     \\
   \centering
   \hspace{-0.9cm} \includegraphics[scale=0.8]{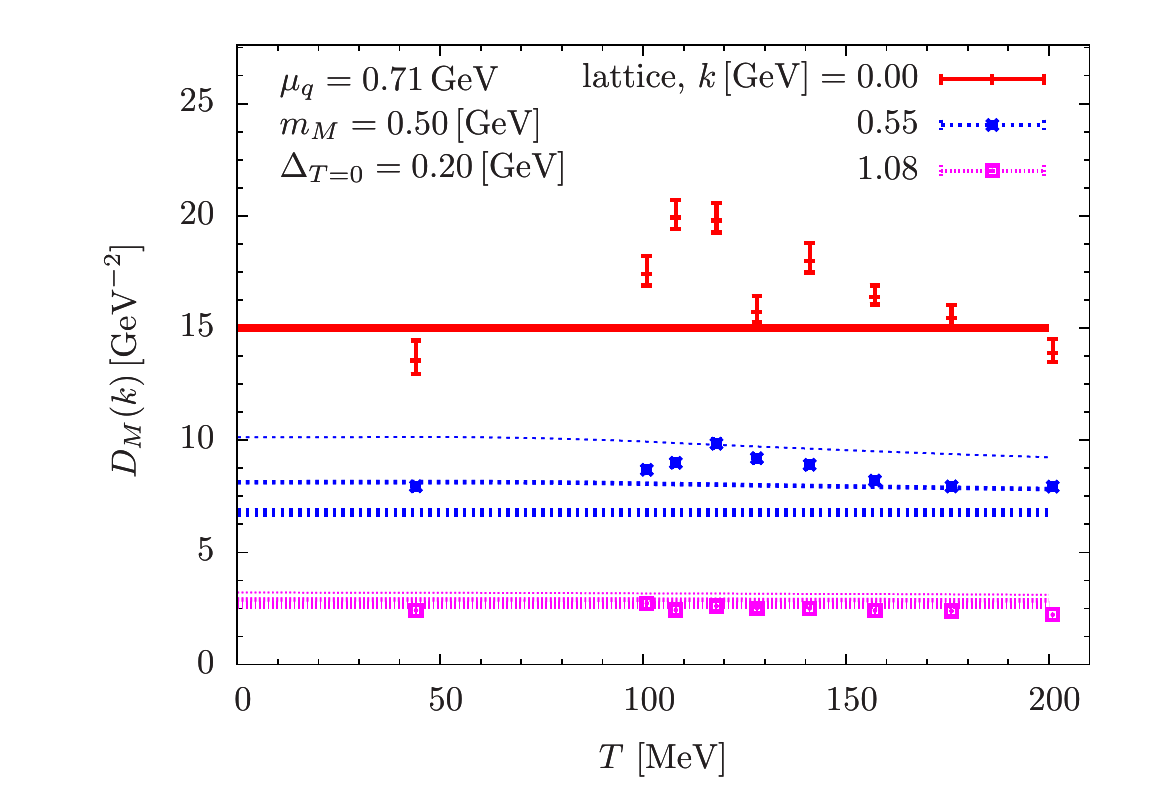} 
   \caption{The $T$-dependence of the gluon propagators at spatial momenta $|\vk|=0.0,\, 0.55,\, 1.08$ GeV and at $\mu_q \simeq 0.71$ GeV. The upper (lower) panel is for electric (magnetic) propagators. The plots with error bars correspond to the lattice data. From bold to thin lines, they correspond to the analytic results at $\alpha_s=$ 0.0, 0.5, 1.0, and the $\Delta_{T=0} =0.2$ GeV.
   }
   \label{fig:mu705T118_fixed_mom}
\end{figure}
Finally we fix $\mu_q=0.71$ GeV and examine the temperature dependence of gluon propagators. The lowest temperature is $T\simeq 44$ MeV, and the next is $\simeq 108$ MeV which is beyond or close to $T_{\rm SF}$. As we have not included the temperature corrections to gluon loops, we stop the comparison around $\simeq$ 150-200 MeV where thermal gluons of densities of $\sim T^3$ should not be ignored. 

For the electric sector, the most notable is the behavior of the softest mode, $D_E (|\vk|=0)$. In analytic results assuming thermal quark excitations, the damping starts to take place around $T\simeq 30$ MeV, and is completed around $T=T_{\rm SF}$ reaching $\sim (g_s p_F)^{-2}$, as expected from usual Debye screening arguments. Meanwhile, the lattice data seems to show slower damping. To adjust the analytic results to the lattice data, we would need gaps about $\Delta_{T=0} \sim$ 300-400 MeV which are perhaps too large. A more reasonable way to reconcile the analytic results with the lattice data is to modify the temperature dependence of $\Delta_T$; if the $\Delta_T$ is stiffer against thermal corrections, i.e., if the abundance thermal quarks are suppressed, then $D_E(|\vk|=0)$ is more stable at low temperature and more quickly melt near $T_{\rm SF}$ than in the present result. 

These considerations bring us to the conjecture that thermal excitations are color-singlet hadrons. If true, thermal loops coupled to external gluon lines are those of hadrons, rather than of quarks. Accordingly the availability of thermal particles are controlled by the Boltzmann factor is $\sim \rme^{-2 \Delta_T/T}$, rather than $\sim \rme^{-\Delta_T/T}$, 
because hadrons contain at least two excited quarks out of condensates.
At the beginning hadrons form a dilute gas, and with increasing $T$ those hadrons overlap allowing thermal quark descriptions. We expect this starts to take place at $T < T_{\rm SF}$ gradually.
As for the magnetic sector, the softest mode, $D_M (|\vk|=0)$, shows slight enhancement around $T\simeq T_{\rm SF}$, but the overall behavior is rather insensitive to $T$.

\section{Summary and discussion}
\label{sec:summary}

We have examined the Landau gauge gluon propagators at finite density and temperature. With analytic one-loop expressions added to massive vacuum gluon propagators, we compare the analytic results with the lattice data in the domains (i) $T\simeq 44$ MeV for $\mu_q=0$-$1.2$ GeV; (ii) $T\simeq 118$ MeV for $\mu_q=0$-$1.2$ GeV; and (iii) $T\simeq $ 0-200 MeV for $\mu_q=0.71$ GeV, for which the lattice data are available for several spatial momenta.  
The gap and critical temperature of the diquark condensation is assumed to be $\Delta_{T=0} = 200$ MeV and $T_{\rm SF} \simeq 114$ MeV, respectively.

The behaviors beyond $T_{\rm SF}$ are overall consistent with the standard descriptions based on the thermal quark loops. The electric sector acquires the Debye mass $\sim g_s p_F$ from gapless modes. In the magnetic sector the paramagnetic behaviors at finite momenta are in agreement between the analytic one-loop results and the lattice's. But the analytic results do not describe slight enhancement in the static mode of the lattice results.

At low temperature $T\simeq 44$ MeV the lattice data shows that the electric and magnetic propagators show only rather weak dependence on the medium effects. To reproduce the behavior in the electric sector it is essential to include the diquark gap, otherwise the Debye screening strongly changes the analytic results. In the previous study we found that $\Delta_{T=0} \simeq 200$ MeV gives the results consistent with the lattice's, while the choice $\Delta_{T=0} \simeq 100$ MeV is a bit too small to reproduce the data at finite momenta. In contrast, the magnetic sector in the analytic results remains insensitive to details of $\Delta_T$.

There are two significant features in the lattice results which cannot be understood in simple terms. The first is the magnetic propagators in the IR limit; at low temperature the magnetic propagators seem slightly weakened as density increases, while at high temperature slightly get enhanced. One solution would be the finite volume artifacts whose impacts should be strongest for soft modes. If not, we need treatments beyond one-loop and/or non-perturbative framework.

The second is the low temperature dependence of the electric propagator at high density, $\mu_q \simeq 0.71$ GeV. In our calculations thermal quarks are gapless and contribute to the Debye screening already at $T\simeq 44$ MeV,
and the screening is very strong at $T\simeq 108$ MeV (the second lowest temperature in Fig.\ref{fig:mu705T118_fixed_mom}). The electric screening at $T\simeq 108$ MeV seems substantially stronger than the lattice data.
 Again the discrepancy can be finite volume artifacts, but in that case the lattice result must be reduced considerably, by $\sim 30$-$40\%$. 
 If this discrepancy remains true for simulations at a bigger volume, 
it will raise questions on the nature of thermal excitations in high density matter, i.e., whether thermal excitations below $T_{\rm SF}$ are thermal quarks or thermal hadrons in superfluid quark matter.

The consequence of thermal excitations are reflected in quantities such as entropies and Polyakov loops. At low temperature the entropy of a hadron gas is much smaller than of a thermal quark gas,
\beq
s_{\rm H}/s_{\rm Q} \sim \rme^{-\Delta_T /T} \,, 
\eeq
as hadronic excitations must contain more than one quark and its energy cost is at least $\sim \Delta_T$. 
This low temperature regime continues, as in a Hagedorn gas, until the Boltzmann suppression is compensated by entropic effects, i.e., drastic growth in the number of hadronic excitations and subsequent overlap of hadrons. 
This Hagedorn type description can give a reasonable description of color-deconfinement at large density, in the same spirit as the transition from a hadron resonance gas to a quark-gluon-plasma at $\mu_q=0$. Below such temperature the matter can be regarded as a quarkyonic matter.

As the entropy is the main ingredient here, the color-deconfinement defined here may occur at $T \lesssim T_{\rm SF}$ (see also Fig.10 in Ref.\cite{Cotter:2012mb} and Fig.2 in Ref.\cite{Iida:2019rah} based on the Polyakov loop). 
Combining this tendency with the insensitivity of $T_{\rm SF}$, we conjecture that the diquark gap is created not through the soft momentum exchange which should be sensitive to medium, but semi-soft or semi-hard momentum transfers which are more robust to the medium effects. We guess this observation for the two-color QCD may be transferred to the three-color case.

Finally we briefly comment on disagreement between gluon propagators based on lattices with coarser grids but bigger volumes \cite{Boz:2018crd}  (which we used in this paper) and those based on finer grids but smaller volumes \cite{Bornyakov:2020kyz}. The latter simulations were done for $N_t \times N_s^3 = (32, 24, 16, 8) \times 32^3$ which are presented as $(0, 188, 280, 560)$ MeV and they found the vanishing of the string tension at $\mu_q \simeq 750$ MeV at the lowest temperature. But if we apply the same estimate on temperature based on $T=1/L_t$, then their results at the lowest temperature are interpreted as the results at $T\simeq140$ MeV. As we discussed in this paper, gluon propagators at $T \gtrsim T_{\rm SF}$ basically follow the explanations based on a thermal quark gas. At such high temperature the sensitivity of various quantities to $\mu_q$ is reasonable, and we regard that the results of Refs.\cite{Boz:2018crd} and \cite{Bornyakov:2020kyz} are rather consistent.

\acknowledgments

We thank the organizers of the workshop ``{\it Probing the physics of high-density and low-temperature matter with ab initio calculations in 2-color QCD}'' which was held in November 2020 at Yukawa Institute of Theoretical Physics, Kyoto University, where this work was motivated.
We are grateful to Profs. A. Maas and J. Skullerud for kindly providing us with their lattice data in Ref.\cite{Boz:2018crd} and explanations concerning with finite volume effects.
T.K. is supported by NSFC grant No. 11875144.


\end{document}